# La Argentina Potencia Atómica

A 70 años del anuncio más espectacular: un sueño, una desilusión y los orígenes del desarrollo nuclear nacional

**Guillermo Jorge**

Universidad Nacional de General Sarmiento

CONICET

Marzo de 2021

gjorge@campus.ungs.edu.ar

"El 16 de Febrero de 1951, en la planta piloto de energía atómica en la isla Huemul, de San Carlos de Bariloche, se llevaron a cabo reacciones termonucleares bajo condiciones de control en escala técnica". Así anunciaba el Gral. Perón, el día 24 de Marzo de 1951, el asombroso resultado de lo que se conoció como Proyecto Huemul. En este artículo repasamos esa historia de desventuras que, a pesar de su extravagante desarrollo y anunciado final, vino a significar el inicio del desarrollo nuclear argentino.

## El anuncio

El día 24 de Marzo de 1951 fue un día excepcional para la historia argentina. El presidente Perón anunciaba en conferencia de prensa, que Argentina tenía el control de la energía atómica con fines pacíficos. El anuncio conmovió al mundo científico de entonces, y también al ambiente político y social local. El comunicado que leyó Perón versaba: "El 16 de Febrero de 1951, en la planta piloto de energía atómica en la isla Huemul, de San Carlos de Bariloche, se llevaron a cabo reacciones termonucleares bajo condiciones de control en escala técnica".

El general Perón continuaba: "los ensayos previos fueron coronado por el éxito, lo que nos alentó para instalar en la isla Huemul una planta piloto de energía atómica con el fin operativo de crear nuevas condiciones de trabajo que permitan la realización total de nuestro proyecto. Así, en oposición con los proyectos extranjeros, los técnicos argentinos trabajaron sobre la base de reacciones termonucleares que son idénticas a aquellas por medio de las cuales se libera la energía atómica en el sol".

El impacto del anuncio en los medios nacionales e internacionales fue amplísimo. *Clarín* titula "La liberación de la energía atómica se logró en el país". *El Orden*: "Nuestro país produce energía atómica, reveló el Gral. Perón". *Noticias Gráficas*: "Provocó sensación el anuncio de que el país tiene la atómica". *La Razón*: "La Argentina ha logrado producir energía atómica".

En el ámbito internacional también se siguió el anuncio con especial atención. *The New York Times* comenta: "El presidente Juan D. Perón anunció hoy que Argentina ha encontrado una manera de producir energía atómica mediante un método nuevo, mucho más simple y más barato que el de cualquier otro país que trabaje en el problema". *The Times* de Londres: "Energía atómica barata a través de un proceso original, según el presidente Perón". Hasta se vieron titulares absurdos, como el del *Jornal do Brasil,* que afirmaba: "El presidente Perón anuncia la explosión de la primera bomba de hidrógeno en Argentina". De todos modos, se puede ver el grado de interés que generó el anuncio tanto en el ámbito local como en el internacional.

El hacedor del importante avance fue un científico de origen austríaco, Ronald Richter, al cual Perón presentó como ciudadano argentino, ya que le había sido otorgada la ciudadanía hacía poco tiempo. Pero, ¿Cómo es que este científico vino a para a la Argentina? y más aún: ¿Qué diablos significaba este anuncio?

## La fisión y la fusión

Detengámonos aquí un segundo. ¿Qué es precisamente lo que se estaba anunciando en la Casa Rosada? Nada menos la obtención de reacciones de fusión nuclear controladas. No sólo algo que nunca nadie antes había logrado, sino algo que nadie lograría con certeza hasta varios años después. Veamos qué es esto de la fusión y la fisión nuclear.

El proceso de *fisión* nuclear ocurre cuando un neutrón (partícula subatómica neutra que integra un núcleo atómico junto a los protones) es bombardeado contra un átomo grande (usualmente suele ser uranio o plutonio), haciendo que éste se divida en dos átomos más pequeños y liberando altas cantidades de energía en este proceso. A su vez, también se liberan otros neutrones que continúan la reacción en cadena. Las bombas atómicas que se lanzaron en Hiroshima y Nagasaki funcionaban con este mecanismo, al igual que todas las centrales y reactores nucleares operativos.

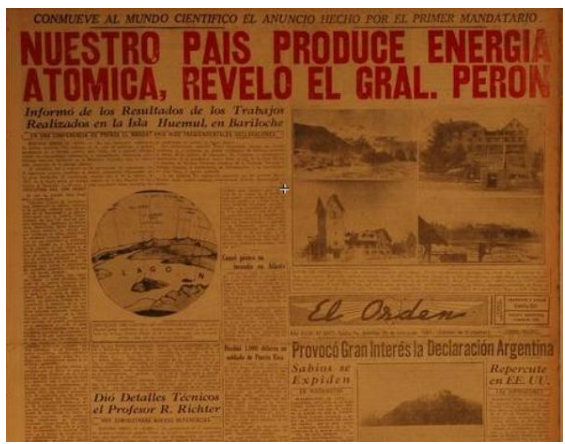
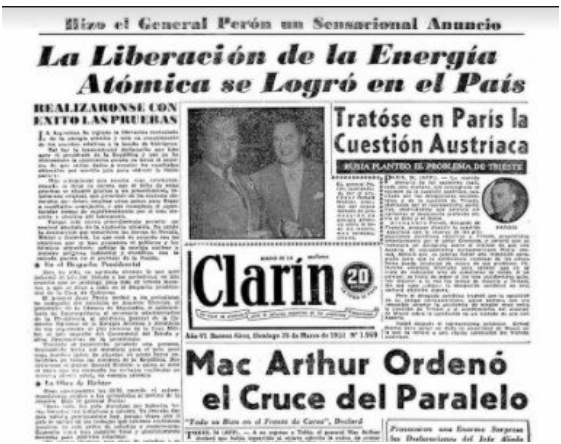

Tapas del diario santafesino *El Orden* y del diario porteño *Clarín*, del día 25 de Marzo de 1951.

Por el contrario, la *fusión* nuclear ocurre cuando elementos livianos son sometidos a una temperatura muy elevada. En estas condiciones, dos átomos livianos pueden chocar y fusionarse en un átomo más pesado, liberando energía en el proceso (por ejemplo, dos átomos de hidrógeno forman uno de helio). El problema reside en que se necesita una temperatura altísima para que esto ocurra, de hecho son los procesos que ocurren dentro de las estrellas. La energía que libera este proceso supera ampliamente la energía obtenida por fisión nuclear. El desafío tecnológico entonces es el de lograr una fusión controlada, y de lograrse sería la energía más barata y limpia de la que dispondría el ser humano, ya que no produciría desechos radiactivos como en el caso de la fisión.

Para posicionarnos en la época, hay que tener en cuenta que la primera explosión de una bomba de hidrógeno (que utiliza el principio de la fusión nuclear) se produjo el 1ro de Noviembre de 1952 en el Atolón de Enewetak (océano pacífico), mientras que la primera reacción termonuclear de fusión controlada tuvo lugar en el Laboratorio Nacional de Los Álamos (Nuevo México, Estados Unidos) recién en el año 1958. Aún hoy, las reacciones termonucleares controladas, con temperaturas superiores a la del sol, sólo pueden ser mantenidas por decenas de segundos, existiendo hasta la fecha algunos proyectos experimentales (como el ITER o el KSTAR), pero ningún reactor operativo que funcione con esta tecnología.

En cuanto a la fisión nuclear, la primera manifestación de su poder incontrolado fue la bomba atómica sobre Hiroshima el 6 de Agosto de 1945, apenas seis años antes del anuncio argentino, y recién en Diciembre de 1951 entró en operación la primera central nuclear de producción de energía. Es decir que para el momento en que se anunció la obtención de fusión nuclear controlada en Argentina, todavía no existían centrales de fisión controlada y menos aún nadie en el mundo había obtenido una reacción controlada de fusión. Es más, hasta el momento del anuncio sólo unos pocos países habían desarrollado reactores nucleares de fisión, y sólo Estados Unidos y la Unión Soviética habían experimentado con bombas atómicas.

## La llegada y el ascenso

Ronald Richter nació en el año 1909 en la ciudad de Falkenau an der Eger, perteneciente entonces al imperio Austro-Húngaro (actualmente Sokolov, República Checa). En 1935 se doctoró en ciencias en la Universidad Carolina de Praga. En el año 1939 se mudó a Alemania, en donde montó un laboratorio financiado por su propio padre en Berlín. Hacia el fin de la guerra, y ante la llegada de las tropas rusas a Berlín, destruyó su laboratorio y huyó a Alemania Occidental.

Luego de su huida, Estados Unidos inició tratativas para contratarlo, pero finalmente estas iniciativas se diluyeron, en palabras de Richter, porque ese país no permitía la entrada de su gato Épsilon. Más allá de esta anécdota, probablemente los Estados Unidos se hayan decidido por otros científicos más reconocidos.

Al año siguiente conoció al ingeniero Kurt Tank (a la postre diseñador del avión argentino Pulqui II), quien sería clave para la llegada de Richter a la Argentina. Tank recuerda que se quedó muy impresionado por la cautivante personalidad del científico. Ese mismo año Kurt Tank es contactado de manera secreta por el Mayor Gallardo Valdéz, que estaba en una misión en Noruega, y por orden directa del gobierno le ofrece un cargo en la fábrica de aviones de Córdoba. Así es como Kurt Tank ingresa clandestinamente a la Argentina con documentos falsos con el seudónimo de Pedro Matthies.

Ya en Argentina, Tank fue consultado sobre posibles técnicos y científicos dispuestos a establecerse en Argentina. Así es como Tank recomienda a Richter. En Mayo del año 1948 se dan los primeros contactos con Richter en París, y finalmente llega a la Argentina el 16 de Agosto de 1948.

A sólo 8 días de estar en el país, Richter concurre a la casa rosada a entrevistarse con Perón. En esa reunión explicó al presidente su plan para generar reacciones nucleares de fusión controlada, con la cual se podría producir energía a un costo ínfimo. Perón sabía que para la concreción de su plan quinquenal iba a necesitar desarrollar nuevas fuentes de energía. No fue difícil, dada la necesidad estratégica y la impronta fascinante del científico, que éste lo convenciera de que el proyecto tenía un alto valor para la República. Perón decide financiar el proyecto y envía a Richter a Córdoba, en donde ya había una comunidad de técnicos alemanes comandada por Kurt Tank, y allí comienza a montar su laboratorio.

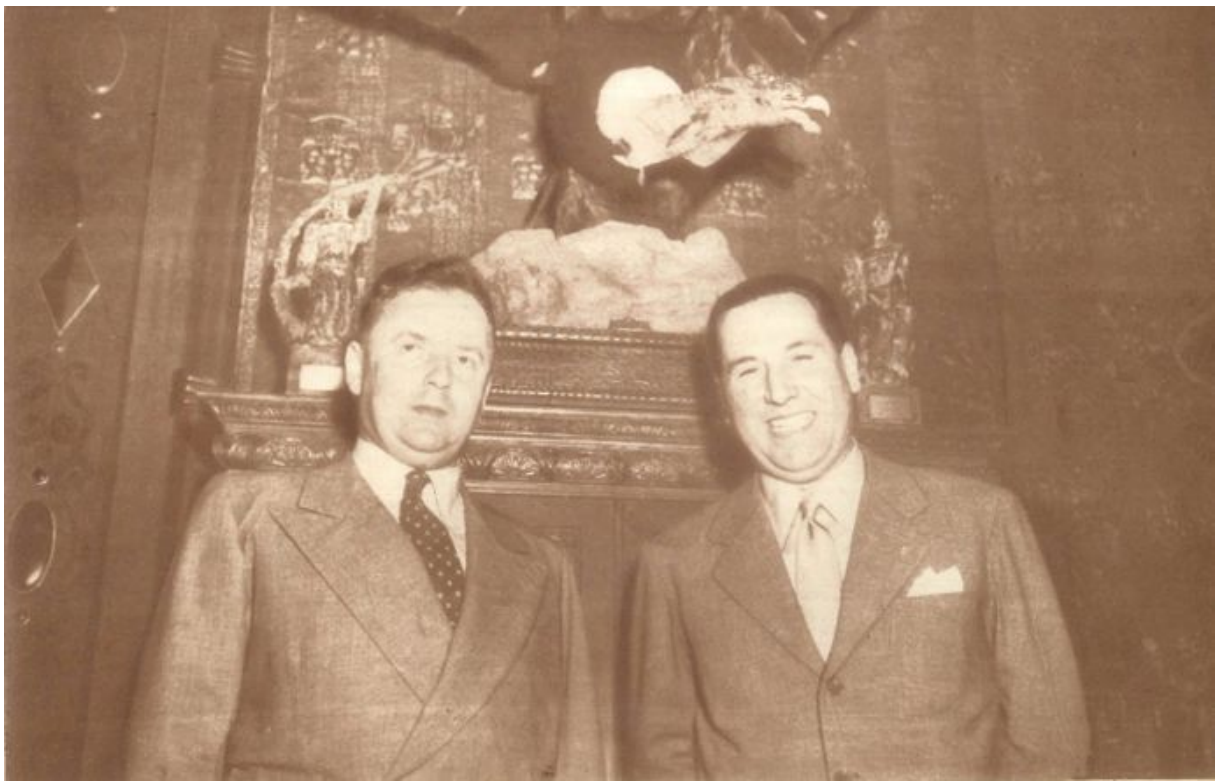

Ronald Richter junto al presidente Juan Perón (Foto: Revista Mundo Atómico, año 1951).

Pero las cosas no fueron demasiado bien al principio. Un incendio en su laboratorio, quizás accidental, lo convence de que estaban intentando sabotear su proyecto. Por eso decide trasladar su laboratorio a un lugar más alejado. Con la ayuda del Coronel Enrique González (por entonces en Migraciones pero que luego sería el primer presidente de la Comisión de Energía Atómica) comienzan a recorrer en avión ciertas zonas de la patagonia. Al sobrevolar Bariloche Richter divisa una isla en el lago Nahuel Huapi, la isla Huemul. Una pequeña isla deshabitada, cerca de la costa del lago, con la ciudad de Bariloche sólo a unos kilómetros. Decide que es el lugar perfecto: discreto, aislado y protegido naturalmente, con disponibilidad ilimitada de agua y en cercanías de una ciudad.

En Junio de 1949 se comienzan los trabajos en la Isla. Y finalmente Richter se muda a Bariloche el 22 de Marzo del año 1950.

### El Proyecto Huemul, la cumbre

Al comenzar el año 1950 el apoyo del presidente Perón al proyecto era total. Se le otorga a Richter la ciudadanía argentina incluso pasando sobre las leyes migratorias de entonces, ya que no cumplía con el tiempo de residencia establecido. El soporte económico era substancial, todo lo que ordenaba Richter se compraba o se construía. De hecho, se construyeron varios laboratorios en la isla, aparte de la usina y una residencia, y el edificio principal del reactor.

El edificio del reactor, de forma cilíndrica, tenía 12 metros de alto por 12 de diámtero y estaba construido con paredes de 1 metro de espesor. Sin embargo, al poco tiempo de haberse terminado y aduciendo fallas en su construcción, Richter mandó a demolerlo. Estas situaciones, sumadas a otras idas y venidas en construcciones y compras, hicieron que González comience a mirar al proyecto con desconfianza.

Luego de la visita del General Perón y su esposa Eva Duarte a la isla, se decide la creación de la Comisión Nacional de Energía Atómica (CNEA), que fue presidida por el propio González. La gran afluencia de recursos económicos y humanos abocados al proyecto hace que la presión aumente sobre Richter para mostrar algún tipo de resultado que demuestre que su idea estaba funcionando. Richter se dedica a la experimentación sobre confinamiento de plasma (materia ionizada a muy alta temperatura) en uno de los laboratorios de la isla.

Es así que el 16 de Febrero de 1951 se produce el ensayo que luego daría lugar al anuncio de prensa. Fue un comienzo de año espectacular para el gobierno. El 9 de Febrero se presentó el Pulqui II, un avión a reacción de avanzada para la época, piloteado por el propio Kurt Tank desde Córdoba hasta Buenos Aires. Por el otro lado, se obtenían reacciones termonucleares el 16 de Febrero, que fueron anunciadas el 24 de Marzo. El 28 de Febrero el presidente le otorga poderes presidenciales a Richter, nombrandolo su representante en la Isla. Cualquier orden que provenga de él en la isla era equivalente a que la diera el propio presidente. El mismo día se le otorga la medalla de honor peronista y se lo nombra Doctor Honoris Causa de la Universidad de Buenos Aires.

Richter había llegado a su cima. Su poder era total y su influencia sobre el presidente fue única. Cualquiera que conociera a Perón daba cuenta de su capacidad de liderazgo y su carisma. Sin embargo, Richter ejercía sobre él algún tipo de encantamiento. La propia personalidad del científico y la idea de Perón de que el proyecto Huemul era la salida a una situación energética cada vez más complicada fueron probablemente las causas de la fascinación que sentía Perón sobre el científico.

Sin embargo, no todos estaban así de encantados. El propio González ya estaba bastante cansado de las idas y vueltas con las compras, las constantes acusaciones de sabotaje o espionaje a distintos miembros del equipo, y hacia mitad de 1951, lo que fue el cambio del grupo de constructores del ejército por una compañía privada. Para colmo, ya con el edificio del reactor demolido, ahora Richter ordenaba la construcción de un reactor bajo tierra. Finalmente hubo otros anuncios de avances durante el año 1951 que mantuvieron con vida el proyecto.

A comienzos de 1952 la situación se volvió insostenible. Ahora Richter comentaba que quería mudar la totalidad del proyecto a otra locación, luego de haber gastado un presupuesto más que importante en las diferentes construcciones y demoliciones y con el edificio del reactor bajo tierra terminado. Ante esta situación, González intenta que Perón conforme una comisión fiscalizadora integrada por científicos y técnicos argentinos para evaluar el devenir del proyecto.

  Convencido Perón de la necesidad de una visita de científicos para evaluar el proyecto, le escribe una carta a Richter para que acepte la visita. La comisión estaría conformada por el Padre Juan Bussolini (director del Observatorio de Física Cósmica de San Miguel), el Ingeniero Manuel Beninson, el Ingeniero Otto Gamba y el Ingeniero Mario Báncora. Por diversas razones no se logra la visita a la isla, sin embargo la comisión recomienda la suspensión temporal del apoyo económico. Una audiencia de Richter con Perón hace que la suspensión del apoyo económico no se produzca, lo cual devino en la renuncia de González a la Comisión de Energía Atómica. Su lugar fue ocupado por Pedro Iraolagoitía.

  Sin embargo Richter siguió desvariando en su conducta a la hora de dirigir el proyecto, y en un cambio de marcha inesperado ordenó que el reactor bajo tierra se rellene completamente y se construya otro al aire libre. Quizás este fue el detonante final. Iraolagoitía, preocupado por los avatares de su antecesor en el cargo, visita la isla. Luego insiste a Perón en la necesidad de la conformación de una comisión fiscalizadora que se entreviste *in situ* con Richter. Sin margen de maniobras, Richter acepta la visita de la comisión, mientras que Perón ordena que todos los equipos comprados por Richter queden en la CNEA.

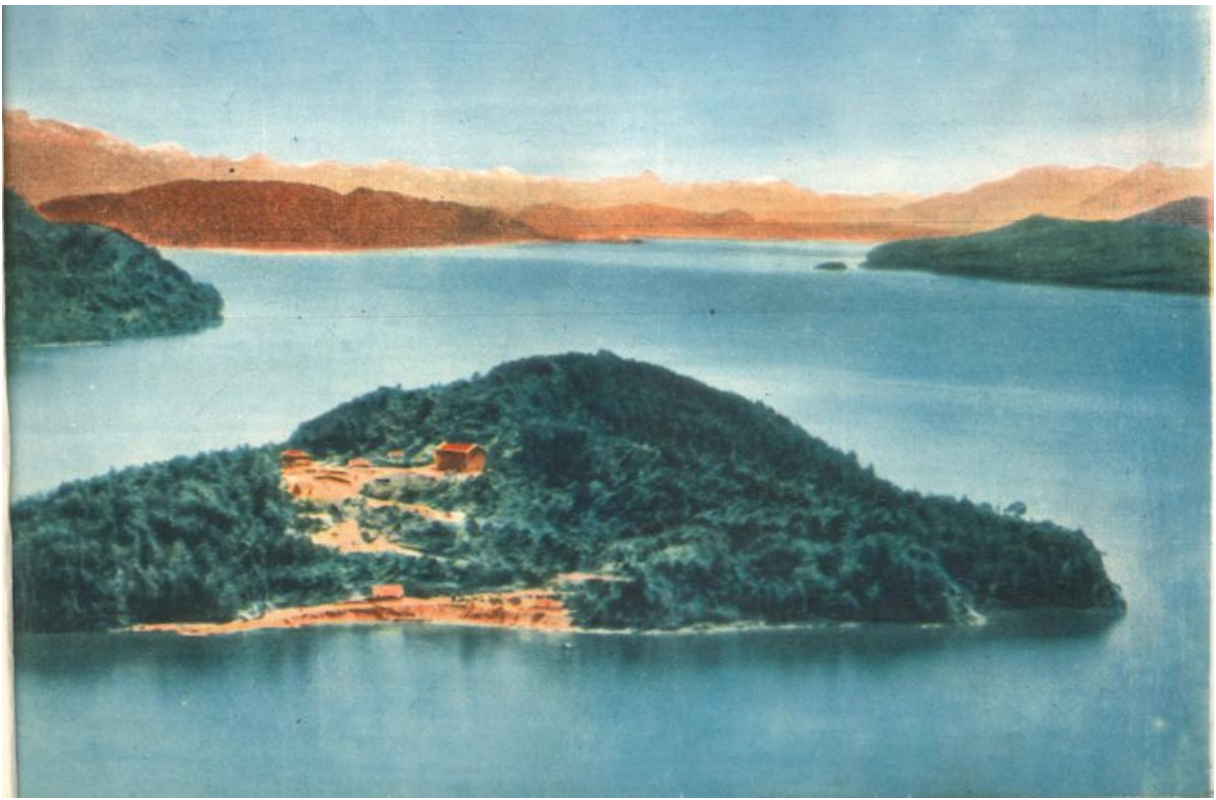

Isla Huemul (Ilustración: Revista Mundo Atómico, año 1951).

### La caída

La segunda comisión fue conformada por Gamba, Beninson, Bussolini y Báncora, a la cual se sumó un joven físico: José Balseiro. Esta comisión visitó la isla junto a un grupo de legisladores nacionales. Allí intentan reproducir los resultados que Richter aducía que provenían de reacciones termonucleares.

El informe de la comisión fue lapidario. No había ningún tipo de indicio científico de que esas observaciones hayan sido causadas realmente por reacciones nucleares. Perón le exige a Richter una respuesta escrita al informe, y ordena a su vez una auditoría sobre el informe y su respuesta a otra comisión de físicos (Ricardo Gans y Antonio Rodríguez). Esta comisión avaló el informe de la comisión fiscalizadora. El futuro de Richter y de su proyecto estaba marcado.

El 24 de Octubre de 1952 se suspenden todos los trabajos en la isla, y finalmente ésta fue intervenida el 22 de noviembre, mientras Richter intentaba entrevistarse con Perón en Buenos Aires, lo cual no logra ya que éste no lo recibe. Finalmente Richter abandona Bariloche para instalarse en un suburbio de Buenos Aires en febrero del año 1953.

### El nacimiento

Con el proyecto dado de baja, había que decidir qué se hacía con el equipamiento y la infraestructura que se había generado. Los equipos fueron trasladados al continente. En cuanto a los edificios de la isla, quedaron en desuso y con el propio paso del tiempo, el avance de la naturaleza y la ayuda de un poco de vandalismo, se convirtieron en ruinas que se pueden observar hasta el día de hoy.

Por suerte la historia no condenó a muerte al centro de desarrollo que se generó en la región. En 1954 y 1955 se realizaron escuelas de física en Bariloche. Esas escuelas fueron el puntapié para la creación del Centro Atómico Bariloche y del Instituto de Física de Bariloche, que quedaron bajo el mando de José Balseiro, el cual daría nombre al famoso instituto luego de su muerte en 1962.

Tras el golpe militar que derrocó a Perón en Septiembre de 1955 se inician investigaciones sobre el manejo científico y económico del Proyecto Huemul. Dichas investigaciones fueron realizadas por una comisión integrada por los físicos Isnardi, Collo y Galloni. También se le retira el doctorado Honoris Causa a Richter, el cual se sume en un ostracismo mantenido hasta su muerte en 1991. Finalmente, se refunda la Comisión Nacional de Energía Atómica en el año 1956.

### Epílogo

Como hemos visto, el recorrido histórico de este estrafalario suceso tuvo algunos resultados muy positivos. A pesar de su costo de 62.428.731 $ moneda nacional, según la investigación de la comisión de Isnardi, que equivaldrían a 31.467.942 U$S a valor 2021. A pesar del oprobio internacional al que quedó expuesto el gobierno argentino. A pesar de haber relegado a la ciencia local sólo a un nivel de fiscalización debido a las diferencias políticas entre la sociedad científica y el gobierno. A pesar de todo, nacieron instituciones de importancia vital y reconocimiento mundial como la Comisión de Energía Atómica, el Centro Atómico Bariloche y el Instituto Balseiro.

Quedan algunas preguntas que quizás nunca serán contestadas. Cómo pudo un científico ignoto, sin conocer el idioma y con costumbres bastante diferentes a las locales, convencer a un hombre de la talla y la inteligencia de Juan Perón sobre su ambicioso proyecto. Por qué ejerció tal poder de convencimiento y fascinación sobre él, quizás el único hombre en el mundo en lograrlo. Por qué le dio Perón tanto poder. Y qué pasó con él durante sus casi cuarenta años de ostracismo en Monte Grande.

Siempre nos quedaremos con estas preguntas, pero mientras tanto podremos recordar aquel momento en el que Argentina fue potencia atómica.

**Para leer más**

Una excelente crónica de los sucesos aquí descritos se encuentra en el libro de Mario Mariscotti, el cual fue la principal fuente de información para este artículo y es considerado como el libro fundamental de este suceso histórico. Los hechos también fueron relatados en numerosos libros, artículos y documentales, con mayor o menor éxito. Entre estos destaco el libro de Matías Alinovi y el de Zulema Marzoratti. Sobre una historia más amplia de la energía nuclear en Argentina puede consultarse el libro de Diego Hurtado.